\documentclass[a4paper,usenatbib,times,fleqn]{mnras}
\usepackage{natbib,graphicx,amsmath,amssymb,times}
\usepackage{aas_macros,url,bm}

\title[Dust-gas mixtures in SPH]{A solution to the overdamping problem when simulating dust-gas mixtures with smoothed particle hydrodynamics}

\author[Price \& Laibe]{Daniel J. Price$^{1}$\thanks{daniel.price@monash.edu} and Guillaume Laibe$^{2}$\thanks{guillaume.laibe@ens-lyon.fr},  \\
$^{1}$School of Physics \& Astronomy, Monash University, Clayton, Vic 3800, Australia \\
$^{2}$Univ Lyon, Univ Lyon1, Ens de Lyon, CNRS, Centre de Recherche Astrophysique de Lyon UMR5574, F-69230, Saint-Genis-Laval, France
}
\pagerange{\pageref{firstpage}--\pageref{lastpage}} \pubyear{2020}

\begin{document}
\label{firstpage}
\bibliographystyle{mnras}
\maketitle

\defcitealias{laibeprice12}{LP12a}
\defcitealias{laibeprice12a}{LP12b}

\begin{abstract}
We present a fix to the overdamping problem found by Laibe \& Price (2012) when simulating strongly coupled dust-gas mixtures using two different sets of particles using smoothed particle hydrodynamics. Our solution is to compute the drag at the barycentre between gas and dust particle pairs when computing the drag force by reconstructing the velocity field, similar to the procedure in Godunov-type solvers. This fixes the overdamping problem at negligible computational cost, but with additional memory required to store velocity derivatives. We employ slope limiters to avoid spurious oscillations at shocks, finding the van Leer Monotonized Central limiter most effective.
 \end{abstract}



\section{Introduction}
\label{sec:intro} 

 In \citet{laibeprice12,laibeprice12a} (hereafter LP12a,b) we found three problems when using Lagrangian particles to simulate the dust component of dust-gas mixtures:  i) artificial trapping of dust particles below the gas resolution, ii) overdamping of waves and slow convergence at high drag, requiring prohibitive spatial resolution iii) timestepping, requiring timesteps shorter than the stopping time, or an implicit scheme \cite[e.g.][]{monaghan97a,miniati10,baistone10,lorenbate14,yangjohansen16,stoyanovskayaetal18,monaghan20}.
 
 In our 2012 study, using smoothed particle hydrodynamics (SPH; \citealt{monaghan92}), we found our numerical solutions for linear waves to be over-damped compared to the analytic solution when the drag between dust and gas was high, i.e., for small grains. \citet{miniati10} similarly found only first order accuracy in the stiff regime when simulating dust as particles and gas on a grid (see also \citealt{yangjohansen16}). This is the `overdamping problem'.
 
 In \citealt{laibeprice14,laibeprice14a} we solved this problem by re-writing the dust/gas equations to describe a single fluid mixture (i.e. as a single set of SPH particles with an evolving dust fraction). This approach avoids the overdamping problem but the mixture approach is only suitable for small grains. \citet{stoyanovskayaetal18} showed that overdamping could be avoided even with dust and gas as particles by interpolating the dust and gas velocities to a common spatial position. Our approach is based on a similar idea.
 
 In this paper we show that the overdamping problem in SPH can be solved by applying ideas from Finite Volume codes, namely reconstruction of the velocity field between pairs of gas and dust particles.
 
\section{Methods}
\subsection{Continuum equations}
Consider a gas and dust mixture represented by two different types of particles. The momentum and energy equations are
\begin{align}
\frac{\partial \bm{v}_{\rm g}}{\partial t} + (\bm{v}_{\rm g} \cdot \nabla) \bm{v}_{\rm g} & = -\frac{\nabla P_{\rm g}}{\rho_{\rm g}} + \frac{K}{\rho_{\rm g}} (\bm{v}_{\rm d} - \bm{v}_{\rm g}), \\
\frac{\partial \bm{v}_{\rm d}}{\partial t} + (\bm{v}_{\rm d} \cdot \nabla) \bm{v}_{\rm d} & = - \frac{K}{\rho_{\rm d}} (\bm{v}_{\rm d} - \bm{v}_{\rm g}), \\
\frac{\partial u_{\rm g}}{\partial t} + (\bm{v}_{\rm g} \cdot \nabla) u_{\rm g} & = -\frac{P_{\rm g}}{\rho_{\rm g}} (\nabla \cdot \bm{v}_{\rm g}) + \frac{K}{\rho_{\rm g}}(\bm{v}_{\rm d} - \bm{v}_{\rm g})^2.
\end{align}

\subsection{SPH equations}
Our SPH algorithm follows LP12a,b in everything except the discrete form of the drag terms. We replace these with
\begin{align}
\left. \frac{{\rm d} \bm{v}_a}{{\rm d} t} \right\vert_{\rm drag} =
& + \nu \sum_i \frac{m_i}{(\rho_i + \rho_a) t^{ai}_{\rm s}} \left(\bm{v}^{*}_{ai}  \cdot \hat{\bm{r}}_{ai} \right)\hat{\bm{r}}_{ai} D_{ai} (h), \\
\left. \frac{{\rm d} \bm{v}_i}{{\rm d} t} \right\vert_{\rm drag} = & - \nu \sum_a \frac{m_a}{(\rho_a + \rho_i) t^{ai}_{\rm s}}  \left(\bm{v}^{*}_{ai}  \cdot \hat{\bm{r}}_{ai} \right)  \hat{\bm{r}}_{ai} D_{ai} (h), \\
\left. \frac{{\rm d} u_a}{{\rm d} t} \right\vert_{\rm drag} = & \nu \sum_i \frac{m_i}{(\rho_a + \rho_i) t^{ai}_{\rm s}}  \left(\bm{v}^{*}_{ai}  \cdot \hat{\bm{r}}_{ai}  \right) \left(\bm{v}_{ai}  \cdot \hat{\bm{r}}_{ai}  \right)  D_{ai} (h), \label{eq:dudtsph}
\end{align}
where the index $a$ refers to gas particles while $i$ refers to dust particles, $\nu$ is the number of dimensions, $\bm{v}_{ai} \equiv \bm{v}_a - \bm{v}_i$,   $\bm{r}_{ai} \equiv \bm{r}_{a} - \bm{r}_i$, $D_{ai}(h)\equiv D(\vert  \bm{r}_{ai} \vert, \max[h_a,h_i])$ is a double-humped kernel \citepalias{laibeprice12}, and the stopping time is defined via
\begin{equation}
t_{\rm s}^{ai} \equiv \frac{\rho_a \rho_i}{K (\rho_a + \rho_i)},
\end{equation}
where density is only computed using neighbours of the same type (i.e. gas density on gas particles and dust density on dust particles). Here we assume $K$ constant, but in general $t_{\rm s}$ may be set according to a physical drag law e.g. Epstein drag. The only difference in our formulation of the drag terms compared to \citetalias{laibeprice12} is that we use a reconstructed velocity for the interaction between particle pairs denoted $\bm{v}^{*}$, rather than the velocity at the position of the particle itself. This improves the estimate of the \textit{local} differential velocity.

\begin{figure*}
   \centering
   \includegraphics[width=\textwidth]{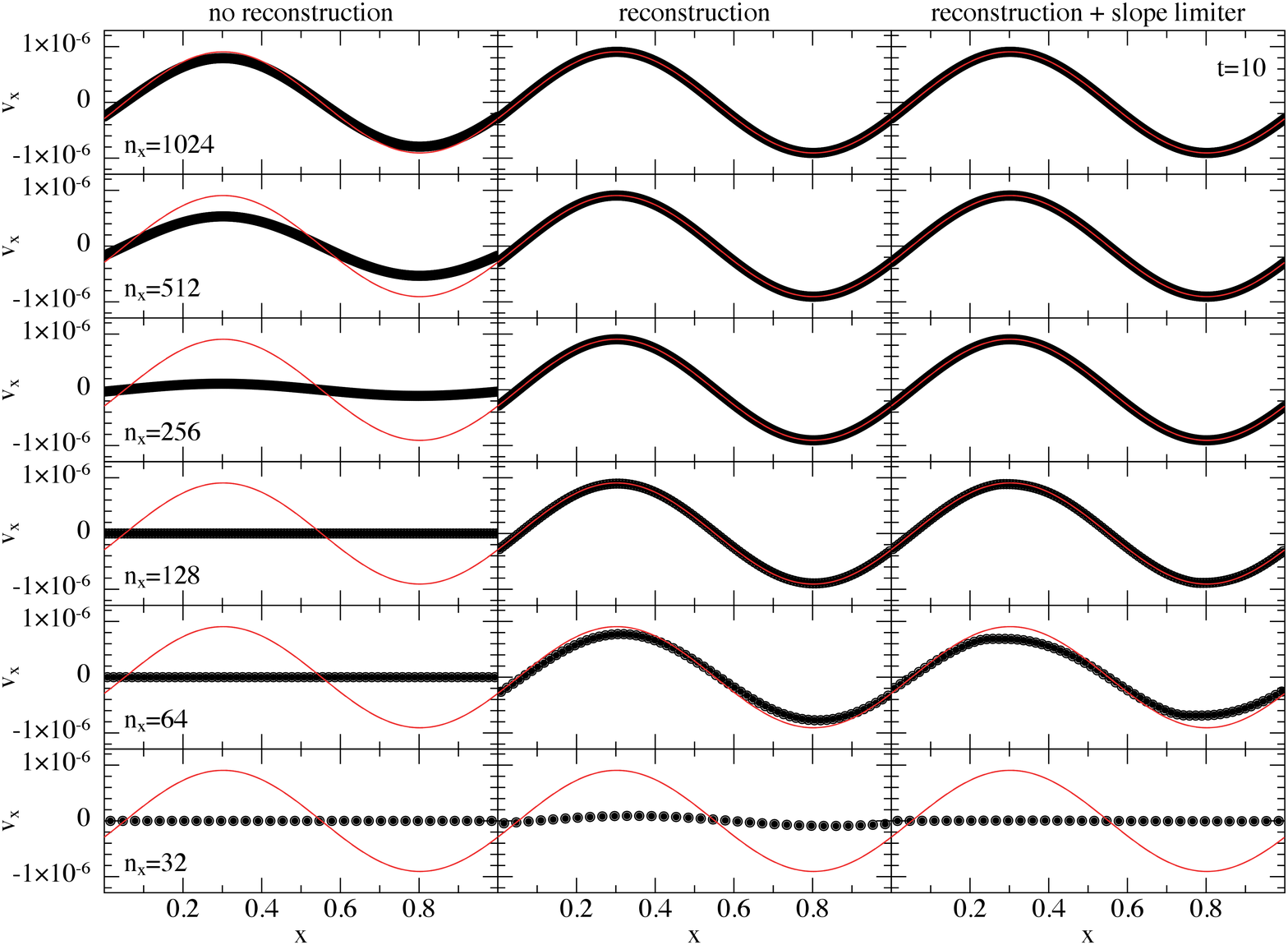} 
   \caption{Dust and gas velocities in the {\sc dustywave} test after 10 wave periods, using $K=1000$ with $2 \times n_x$ particles without reconstruction and with and without the slope limiter (see labels). Reconstruction avoids the need to resolve $h\sim t_{\rm s} c_{\rm s}$ (resolved at $n_x = 1024$ particles). Exact solution shown in red.}
   \label{fig:dustywave}
\end{figure*}


\begin{figure*}
   \centering
   \includegraphics[width=\textwidth]{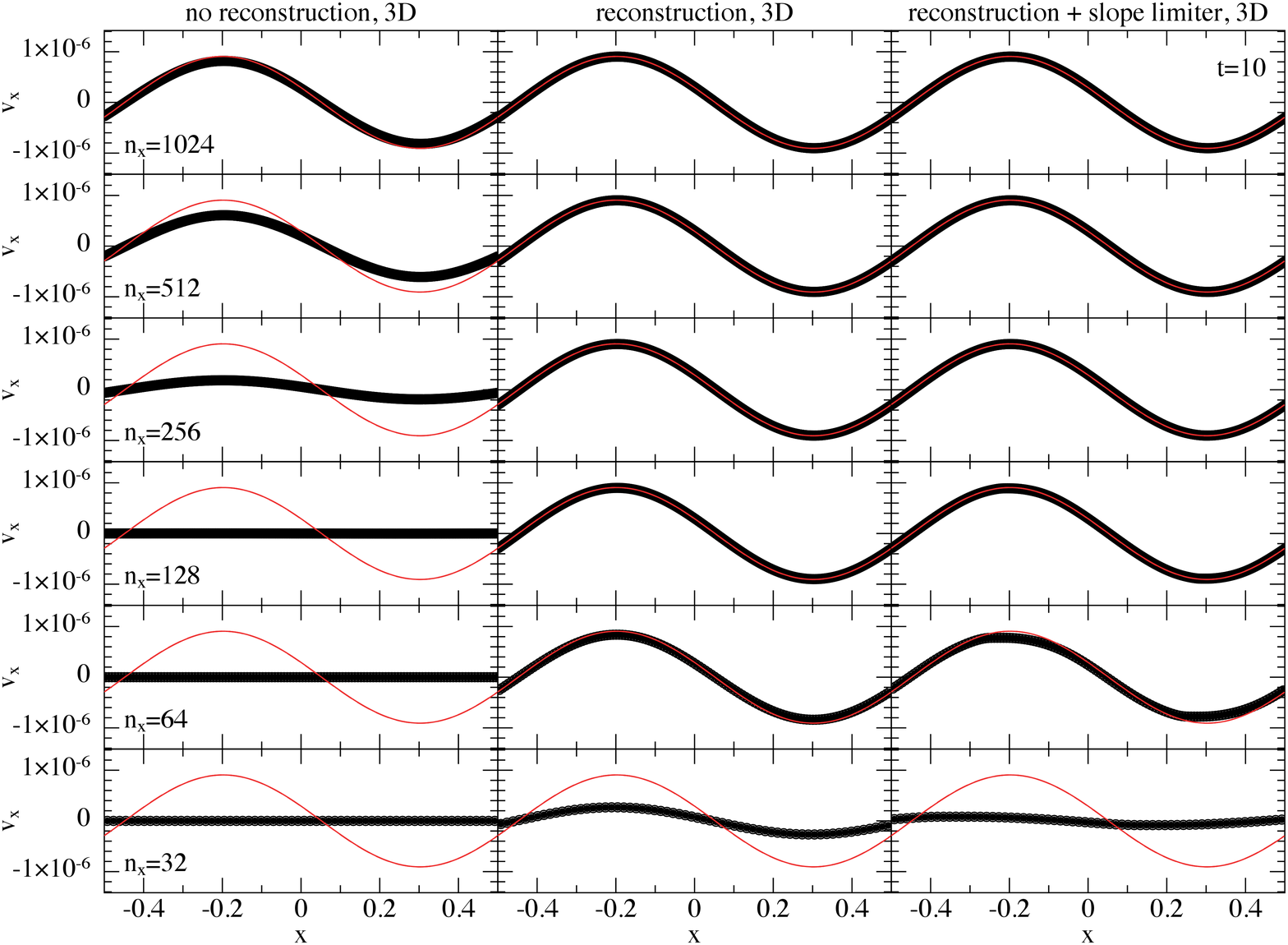} 
   \caption{As in Figure~\ref{fig:dustywave} but in 3D with {\sc phantom} using $n_x \times 12 \times 12$ gas particles (solid) and $n_x \times 12 \times 12$ dust particles (open) initially placed using dense sphere packing. Exact solution from \citet{laibeprice11} shown in red.}
      \label{fig:dustywave3D}
\end{figure*}

\subsection{Reconstruction}
We reconstruct the velocity for each particle pair ($a$,$i$) using 



%
\begin{align}
\bm{v}^{*}_a & = \bm{v}_a + \left(\bm{r}^{*} - \bm{r}_{a}\right)^\beta \frac{\partial \bm{v}_{a}}{\partial \bm{r}_{a}^\beta}; \\
\bm{v}^{*}_i & = \bm{v}_i + \left( \bm{r}^{*} - \bm{r}_{i}\right)^\beta \frac{\partial \bm{v}_{i}}{\partial \bm{r}_{i}^\beta}.
\end{align}
where to avoid confusion with particle labels we use $\alpha$, $\beta$ and $\gamma$ to refer to tensor indices, with repeated tensor indices implying summation. At the barycentre between the particles $a$ and $i$ --- i.e., at $\bm{r}^{*} = \bm{r}_{a} + \mu_{ai} \bm{r}_{ai} = \bm{r}_i - \mu_{ia} \bm{r}_{ai}$, these relations combine to
\begin{equation}
\bm{v}^{*}_{ai} \cdot \hat{\bm{r}}_{ai} = \bm{v}_{ai} \cdot \hat{\bm{r}}_{ai} - \mu_{ai} \vert r_{ai} \vert  \left( S_{ai} + S_{ia} \right) ,
\end{equation}
where $S_{ai} \equiv \hat{r}_{ai}^\alpha \hat{r}_{ai}^\beta \frac{\partial v_{a}^{\alpha}}{\partial x_{a}^\beta} $ and $\mu_{ai} = m_{a} / \left(m_{a} + m_{i} \right)$.
Velocity gradients are computed using an exact linear derivative operator \citep[e.g.][]{price12}, i.e. by solving the $3 \times 3$ matrix equation
\begin{equation}
R_{\beta\gamma} \frac{\partial v^{\alpha}}{\partial r^{\gamma}}  = -\sum_{b}m_{b} v^{\alpha}_{ab}\nabla^{\beta}W_{ab}\left(h_{a} \right),
\label{eq:dvdx}
\end{equation}
where 
\begin{equation}
R_{\beta\gamma} = \sum_b m_b (\bm{r}_b - \bm{r}_a)^\beta \nabla^\gamma W_{ab} (h_a).
\label{eq:rmat}
\end{equation}
 The summations on the right hand side of Equations~\ref{eq:dvdx} and \ref{eq:rmat} are computed during the density summation, with the summation index over particles of the same type. We found no difference using the exact linear operator versus the usual SPH derivative.
\subsection{Slope limiters}
\label{sec:limiters}
The danger with reconstruction is the reintroduction of spurious oscillations when the solution is discontinuous. To prevent this, the factor $\left(S_{ai} + S_{ia}\right)$ may be replaced by a slope limiter, i.e. a function $2 f\left(S_{ai},S_{ia} \right)$ that preserves monotonicity \citep{van-leer74}. We explored a range of limiters (e.g. \citealt{sweby84}) including, from most to least dissipative, minmod
\begin{equation}
f(a, b)  = \begin{cases}
\min(\vert a \vert, \vert b \vert) & a > 0, b > 0 \\
-\min(\vert a \vert, \vert b \vert) & a < 0, b < 0 \\
0 & \text{otherwise},
\end{cases}
\end{equation}
van Leer \citep{van-leer77}
\begin{equation}
f(a, b)  = \begin{cases}
\frac{2 ab}{a + b} & ab > 0 \\
0 & \text{otherwise},
\end{cases}
\end{equation}
van Leer Monotonized Central (MC) \citep{van-leer77}
\begin{equation}
f(a, b)  = \begin{cases}
\text{sgn}(a) \min(\vert \frac12 (a + b) \vert, 2\vert a \vert, 2 \vert b \vert ) & ab > 0 \\
0 & \text{otherwise},
\end{cases}
\label{eq:vanleermc}
\end{equation}
and Superbee \citep{roe86,sweby84}
\begin{equation}
f(a, b)  = \begin{cases}
\text{sgn}(a) \max\left[ \min(\vert b \vert, 2\vert a \vert), \min( 2 \vert b \vert, \vert a \vert ) \right] & ab > 0 \\
0 & \text{otherwise}.
\end{cases}
\end{equation}

\begin{figure}
   \centering
   \includegraphics[width=\columnwidth]{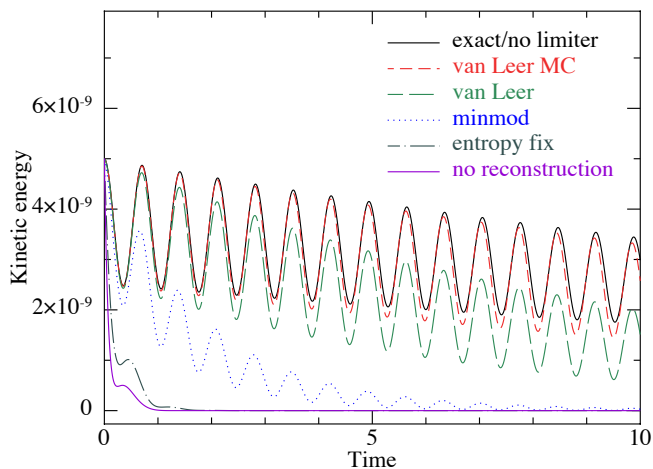} 
   \caption{Kinetic energy as a function of time in the 1D {\sc dustywave} problem, comparing different slope limiters. From top to bottom results employ reconstruction with no limiter, the van Leer MC, van Leer and minmod limiters, our `entropy fix' (Section~\ref{sec:entropy}), and no reconstruction.}
   \label{fig:limiters}
\end{figure}

\begin{figure}
   \centering
   \includegraphics[width=\columnwidth]{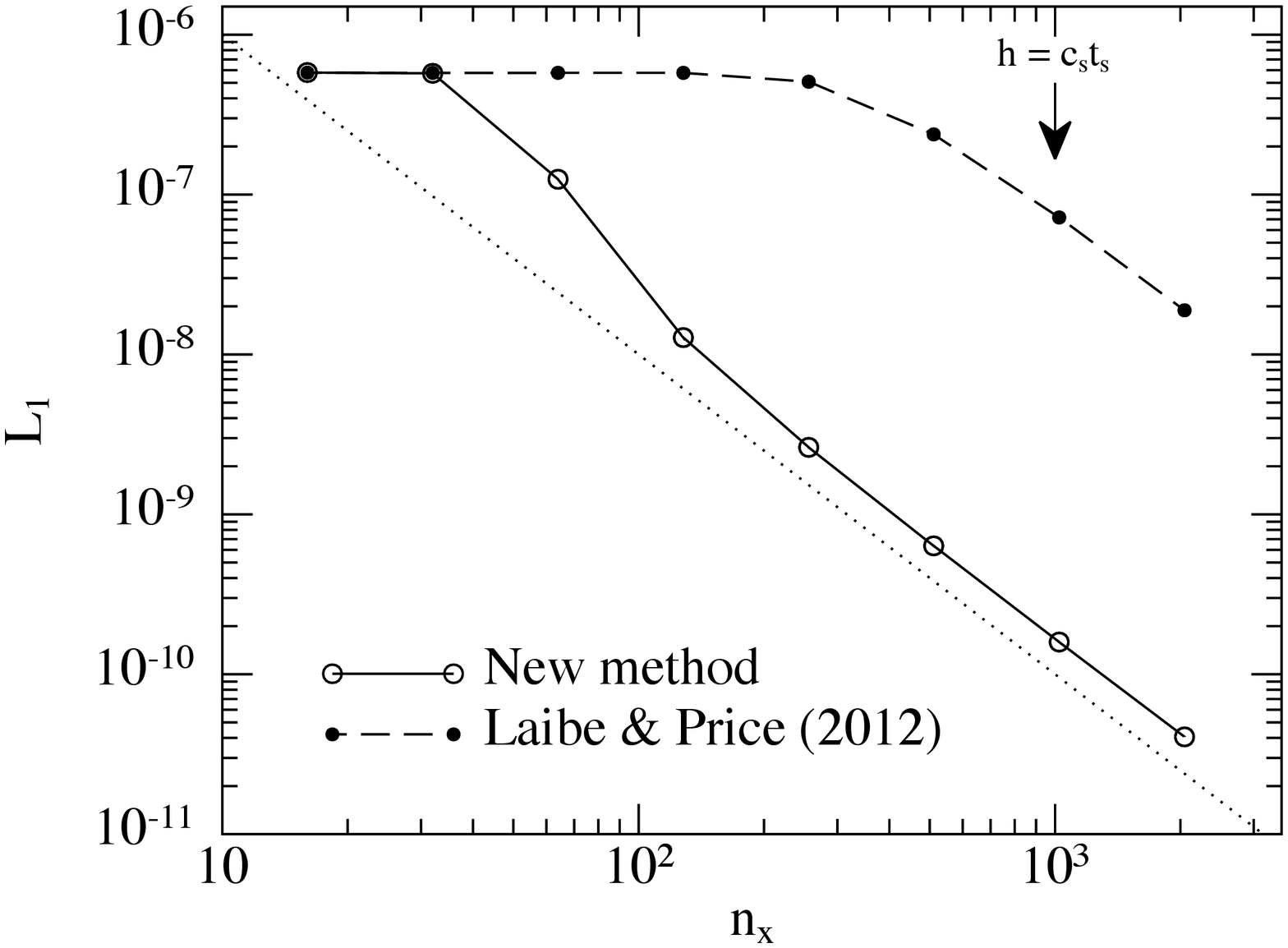} 
   \caption{Convergence on the {\sc dustywave} problem, showing $L_1$ error as a function of the number of particles per wavelength in 1D. Solid line uses reconstruction and the van Leer MC limiter, dashed line no reconstruction. Dotted line shows slope of $-2$ expected for 2nd order. Arrow indicates the no-longer-necessary $h \lesssim c_{\rm s} t_{\rm s}$ criterion required by \citetalias{laibeprice12}.}
      \label{fig:convergence}
\end{figure}

\subsection{Slope limiters and entropy}
\label{sec:entropy}
 Slope limiters are usually employed in the context of Total Variation Diminishing (TVD) schemes \citep{harten83}, but application of the TVD concept beyond 1D or to unstructured/meshfree methods is less clear (e.g. \citealt*{csn17}). A physical interpretation can be seen from our Equation~\ref{eq:dudtsph}. For the drag term to provide a positive definite contribution to the entropy $\bm{v}_{ai}\cdot \hat{\bm{r}}_{ai}$ and $\bm{v}^*_{ai} \cdot \hat{\bm{r}}_{ai}$ must have the same sign, such that ${\rm d}u/{\rm d}t\vert_{\rm drag}$ is positive. Pairwise positivity is not strictly necessary so long as the sum over all neighbours is positive. We tried setting $\bm{v}^*_{ai}\cdot \hat{\bm{r}}_{ai} = \bm{v}_{ai}\cdot \hat{\bm{r}}_{ai}$ if the signs differ, but found this to be more dissipative than using slope limiters (see Figure~\ref{fig:limiters}).  We found the van Leer MC limiter to provide the best compromise between monotonicity and dissipation. 

\section{Results}

We test our improved algorithm in 1D using the {\sc ndspmhd} code \citep{price12} and in 3D using {\sc phantom} \citep{priceetal18a}. We use explicit global timestepping with a leapfrog integrator, the M$_6$ quintic kernel for the SPH terms with the double hump M$_6$ employed for the drag terms \citepalias{laibeprice12}. The results are not sensitive to the choice of kernel provided a double hump kernel is used for the drag. The timestep was set to 0.9 times the minimum stopping time (we found that setting $\Delta t = t_{\rm s}$ exactly as in \citetalias{laibeprice12} could result in instability with reconstruction). We use the van Leer MC limiter unless otherwise specified.

\subsection{\sc Dustywave}
\label{sec:dustywave}


\begin{figure*}
   \centering
   \includegraphics[width=\textwidth]{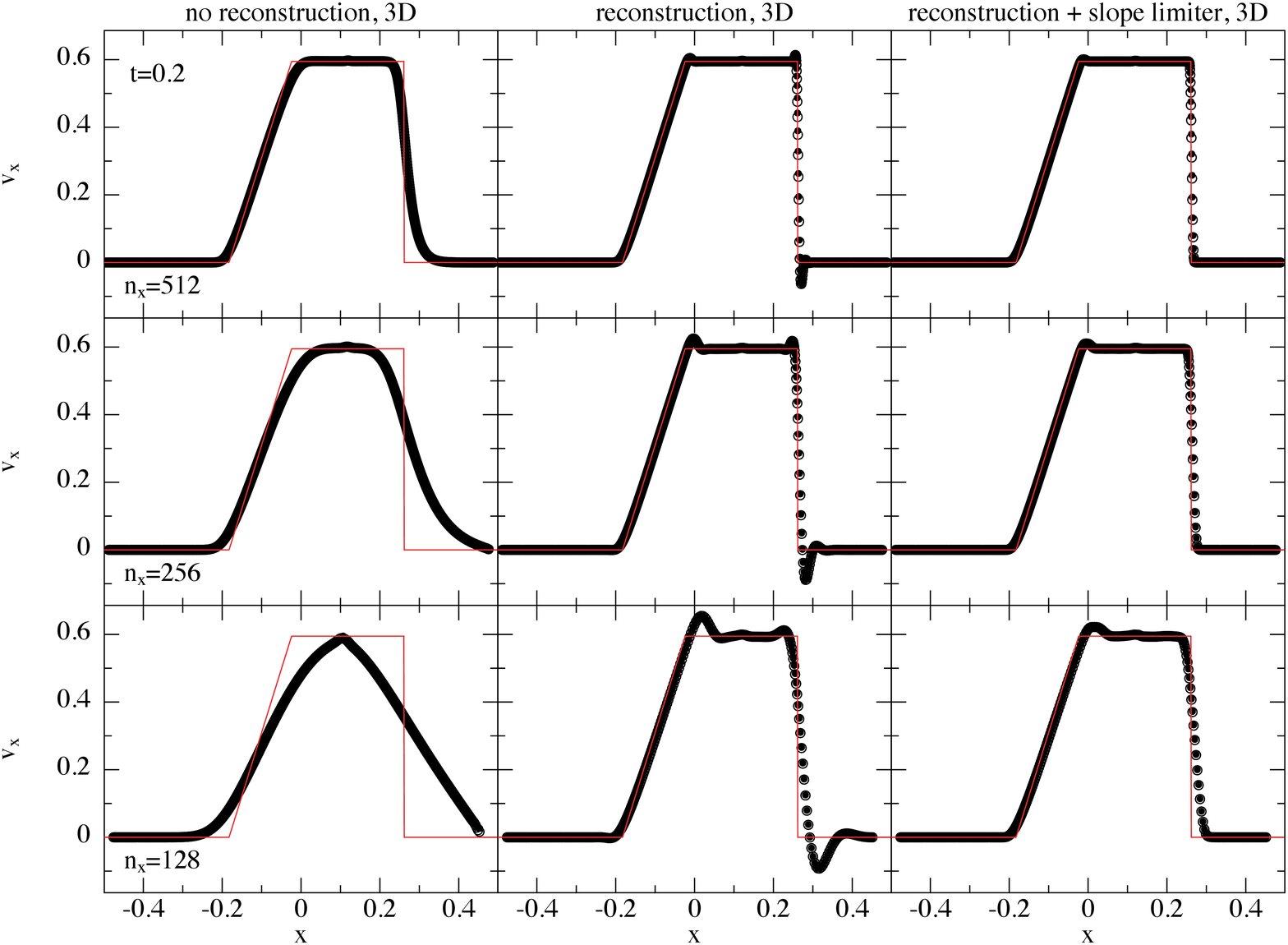} 
   \caption{Results of the {\sc dustyshock} problem performed in 3D with {\sc phantom}, performed at three different numerical resolutions (bottom to top) with no reconstruction (left column), with reconstruction but no slope limiter (middle) and using reconstruction with the van Leer MC limiter (right column). Exact solution in red, points show velocity on gas (solid) and dust (open circles) particles.}
   \label{fig:dustyshock3D}
\end{figure*}

Figure~\ref{fig:dustywave} shows the results of the {\sc dustywave} described in \citet{laibeprice11}, performed using $2\times n_x$ particles with a fixed drag coefficient $K=1000$, $\rho_{\rm g} = \rho_{\rm d} = 1$ and $c_{\rm s} = 1$ (giving $t_{\rm s} = 5 \times 10^{-4}$) and a perturbation amplitude of $10^{-6}$. We use an adiabatic equation of state $P = (\gamma - 1)\rho u$ with $\gamma = 5/3$ in the gas. In the absence of reconstruction, overdamping occurs when $h \gtrsim c_{\rm s} t_{\rm s}$, i.e. for $n_x \lesssim 1024$ (left column), as found by \citetalias{laibeprice12}. Adding reconstruction captures the true solution to within a few percent for $n_x \gtrsim 64$ (middle column), while the slope limiter does not visibly degrade it (right column).

Figure~\ref{fig:dustywave3D} shows the results in 3D using {\sc phantom}. We follow the procedure used in \citet{priceetal18a}, placing the particles using dense sphere packing and cropping the grid in the $y$ and $z$ directions at 12 particle spacings (for efficiency), giving $2 \times 128 \times 12 \times 12$ particles. The results in 3D are indistinguishable from those shown in Figure~\ref{fig:dustywave}, showing our method also works in three dimensions.

\subsubsection{Choice of slope limiter}
Figure~\ref{fig:limiters} shows the kinetic energy as a function of time in the 1D {\sc dustywave} problem at a resolution of $n_x=128$. The solution with reconstruction but no slope limiter (solid black line) is indistinguishable from the analytic damping rate \citep{laibeprice11}. By contrast, the solution with no reconstruction (magenta line) is damped in less than one wave period. All limiters apart from Superbee (not shown) give results intermediate between these two extremes. Superbee, defined as the least dissipative limiter to satisfy the TVD property \citep{sweby84}, was found to \emph{increase} rather than decrease the kinetic energy and produce a clipped wavefront. This numerical `over-steepening' is a known problem with Superbee \citep[e.g.][]{kleeetal17}. The Van Leer MC limiter gives the closest match to the analytic damping rate while still remaining effective at shocks (see below). More dissipative limiters all bring back some degree of overdamping. No limiter apart from our entropy fix was found to guarantee positive entropy.

\subsubsection{Convergence}
 Figure~\ref{fig:convergence} shows the $L_1$ error $(1/N \sum \vert v_x - v_{x, {\rm exact}} \vert)$ as a function of the number of particles per wavelength for the 1D {\sc dustywave} problem. Without reconstruction convergence is flat at low resolution $(n_x \leq 256$) because the wave is almost completely damped, becoming second order only after the $h < c_{\rm s} t_{\rm s}$ criterion is satisfied ($n_x \gtrsim 1000$). With reconstruction and the slope limiter we find second order convergence for $n_x \gtrsim 32$, once the wave is sufficiently resolved for gradients to be accurate.


\subsection{\sc Dustyshock}

Figure~\ref{fig:dustyshock3D} shows the results of the {\sc dustyshock} test from \citetalias{laibeprice12} at three different numerical resolutions (bottom to top). \citet{lehmannwardle18} also proposed a dusty shock test, but their test is for the intermediate regime where the drag is moderate. Here we are interested in the strong drag regime, where the stopping time is negligible.

We set up the problem as usual with gas with $x < 0$ set up with $(\rho,P,v_x) = (1.0,1.0,0.0)$ and gas with $x >= 0$ set up with $(\rho,P,v_x) = (0.125,0.1,0.0)$. We performed the test in both 1D and 3D but only show results from the 3D calculation since, as for the wave test, they are very similar to those obtained in 1D. In 3D we set the particle spacing using $n_x \times n_y \times n_z$ gas particles for $x \in [-0.5,0.0]$, and $n_x/2 \times n_y/2 \times n_z/2$ gas particles in $x \in [0.0,0.5]$ to resolve the 8:1 density contrast without introducing highly anisotropic initial particle distributions. As for the wave test we crop the domain in the $y$ and $z$ directions to match the particle spacing, using $n_y = 24$ and $n_z = 24$. We initialise the dust as copies of the gas particles, assuming a dust-to-gas ratio of unity. We apply artificial viscosity as usual using the modified version of the \citealt{cullendehnen10} switch (see \citealt{priceetal18a} for details).

 Figure~\ref{fig:dustyshock3D} shows results using the default approach (left column), which at low resolution (bottom left panel) produces a solution appropriate for a smaller drag coefficient. Applying reconstruction with no slope limiter (middle column) the numerical solution is much closer to the exact solution (red line), resolves shock discontinuities to within $\sim 3h$, but produces an unphysical oscillation ahead of the shock front. The right column shows that the slope limiter eliminates such oscillations.The remaining defects in the solution (e.g. at $x=-0.02$) can be seen to disappear as the numerical resolution is increased (right column, bottom to top), with the corresponding $L_1$ error reducing from $1.4 \times 10^{-2}$ at $n_x = 128$ to $6.6 \times 10^{-3}$ using $n_x = 256$ and $4.0 \times 10^{-3}$ using $n_x = 512$.


 
  
We employed $n_x = 11,255$ particles in 1D to obtain reasonable results on this problem in \citetalias{laibeprice12}!

\section{Discussion}
  
  In this paper we have shown how the overdamping problem can be fixed by evaluating the drag at the barycentre of each dust-gas particle pair. The slow convergence observed by \citetalias{laibeprice12} is caused by the particle separation (of order the resolution length, $h$) being too large to correctly resolve the drag lengthscale $l \sim c_{\rm s} t_{\rm s}$. This is why the issue is absent when simulating the dust and gas as a single fluid mixture \citep{laibeprice14,laibeprice14a}. A similar idea of interpolating the velocities to a common spatial position was also employed by \citet{stoyanovskayaetal18} as part of their implicit scheme, where it was also shown to solve the overdamping problem. We used explicit timestepping and employed slope limiters to avoid introducing unphysical oscillations at shock fronts. \citet{fungmuley19} similarly found reconstruction of the velocity field necessary for accurate drag in their semi-analytic hybrid  (dust as particles, gas on the grid) scheme.

  Solving the overdamping problem does not make the other problems go away. Timestepping is relatively easy to solve, with numerous implicit methods already proposed both in the context of SPH \citep{monaghan97a,laibeprice12a,lorenbate14,lorenbate15,stoyanovskayaetal18,monaghan20} and in Eulerian particle-gas codes \citep[e.g.][]{miniati10,baistone10,yangjohansen16,fungmuley19}. Our work makes these worth implementing, since overdamping remains with implicit time integration (see Figures~6--9 of \citealt{lorenbate14}). That is, although these schemes make calculation of small grain species efficient, in the absence of our fix they remain inaccurate at high drag. \citealt{lorenbate14} showed that the overdamping was not as severe when the dust-to-gas ratio is low, which suggests a modified criterion $h < c_{\rm s} t_{\rm s} / \epsilon$. With reconstruction or interpolation no spatial resolution criterion is necessary, as found by \citet{stoyanovskayaetal18}. 
  
   The artificial trapping problem is harder to solve. A single fluid model with no approximations \citep{laibeprice14} can accurately capture waves and shocks for both small and large grains with no artificial trapping \citep{laibeprice14a,bkp19}. However, a single fluid model fails to capture large grains with significant inertia because the dust velocity field is assumed to be single valued everywhere, meaning that dust particles cannot stream or interpenetrate \citep{laibeprice14a}. The domain of validity is thus reduced in any case to the regime of small grains, where the terminal velocity approximation greatly simplifies matters \citep{laibeprice14,pricelaibe15,ballabioetal18}. The single fluid method has been extended to multiple grain species \citep{hpl18,bkp19,lcl19}. But for large grains one is forced to use particles. Our approach to avoid artificial trapping to date has been to over-resolve the gas compared to the dust \citep[e.g.][]{mpp19}. This works but is not fail-safe. Artificial trapping also occurs with tracer particles in Eulerian simulations \citep[e.g.][]{pricefederrath10}, where \citet{cdp19} proposed the `Monte Carlo tracer particle' method as a solution. Whether or not similar ideas could be applied to dust-gas mixtures would be worth investigating.
  
  An obvious extension of our method is to apply the same principles to shock capturing in SPH, by using reconstruction in the artificial viscosity terms. We have published preliminary experiments in a conference proceedings \citep{price19}.  \citet{rosswog19a} has also recently proposed a similar method, using both first and second derivatives in the reconstruction.
  
  The main caveat, which would also apply to shock capturing, is that the entropy increase is not guaranteed to be positive definite. While we found the errors to be small, it would be desirable to guarantee positivity while eliminating overdamping.




\section{Conclusions} 
 We have shown how the overdamping problem when simulating dust-gas mixtures with separate sets of particles in SPH can be solved by `reconstructing' the velocity field between pairs of dust and gas particles using an approach similar to that employed in finite volume schemes. A slope limiter is needed to avoid oscillations at shocks. The advantange of the new method is that the overdamping problem can be solved with minor changes to existing dust-gas SPH codes at negligible computational expense. The disadvantages are that performing reconstruction requires storage of nine velocity derivatives per particle and does not always guarantee positive entropy despite our use of slope limiters. Our algorithm is implemented in the public {\sc phantom} code \citep{priceetal18a}.
  
\section*{Acknowledgments}
We thank Pablo Loren-Aguilar, Matthew Bate, Christophe Pinte, Ugo Lebreuilly, Benoit Commer\c{c}on, Jim Stone and James Wadsley for useful discussions, and the referee for helpful comments. DP thanks Bernhard Mueller for useful lecture notes, his teaching load for inspiration, and is grateful for funding from the Australian Research Council via FT130100034 and DP180104235. We acknowledge computing time on Gadi via the Australian National Compute facility, and on Ozstar, funded by the Australian Government and Swinburne University. GL acknowledges funding from PNP, PNPS, PCMI of CNRS/INSU, CEA and CNES, France, and via the IDEXLyon project (contract ANR-16-IDEX- 0005) under Univ. Lyon. We acknowledge European Research Council (ERC) funding under H2020 grant 864965. We used {\sc splash} \citep{price07}.

\bibliography{dan}

\label{lastpage}
\end{document}